\documentstyle[aas2pp4]{article}
\input epsf
\begin{document}
 
\rightline{UIUC-THC-98/2}
 
\title{Complementary Measures of the Mass Density and Cosmological Constant}
\author{Martin White}
\affil{Departments of Astronomy and Physics\\
University of Illinois at Urbana-Champaign\\
Urbana, IL 61801}
\authoremail{white@physics.uiuc.edu}
  
\begin{abstract}
\noindent
\rightskip=0pt
The distance-redshift relation depends on the amount of matter of each type
in the universe.  Measurements at different redshifts constrain differing
combinations of these matter densities and thus may be used in combination
to constrain each separately.
The combination of $\Omega_0$ and $\Omega_\Lambda$ measured in supernovae at
$z\la 1$ is almost orthogonal to the combination probed by the location of
features in the cosmic microwave background (CMB) anisotropy spectrum.
We analyze the current combined data set in this framework, showing that the
regions preferred by the Supernova and CMB measurements are compatible.
We quantify the favoured region.
We also discuss models in which the matter density in the universe is
augmented by a smooth component to give critical density.
These models, by construction and in contrast, are not strongly constrained
by the combination of the data sets.
\end{abstract}
 

%\twocolumn

\section{Introduction}

There has recently been progress in observationally determining the
cosmological redshift-distance relations, which in an FRW universe
determine the contribution of various species to the total energy density.
Two methods in particular have advanced rapidly: the measurement of Type-Ia
supernovae (SN-Ia) as ``standard'' or ``calibrated'' candles at high
redshift, and the angular size of features in the cosmic microwave background
(CMB) temperature anisotropy power spectrum.

At low-$z$ the redshift-distance relation determines the deceleration parameter
$q_0$, however as one goes to higher $z$ the combination of densities which
is constrained changes (e.g.~Goobar \& Perlmutter~\cite{GooPer}).
At extremely high-$z$, as probed by the CMB anisotropies, the combination of
densities which is probed can be nearly orthogonal to the combination given
by $q_0$ (White \& Scott~\cite{WhiSco}).  Thus a combination of these two
measurement techniques can isolate a small region in parameter space much
more effectively than either of these methods independently.

\begin{figure}[t]
\begin{center}
\leavevmode
\epsfxsize=3.4in \epsfbox{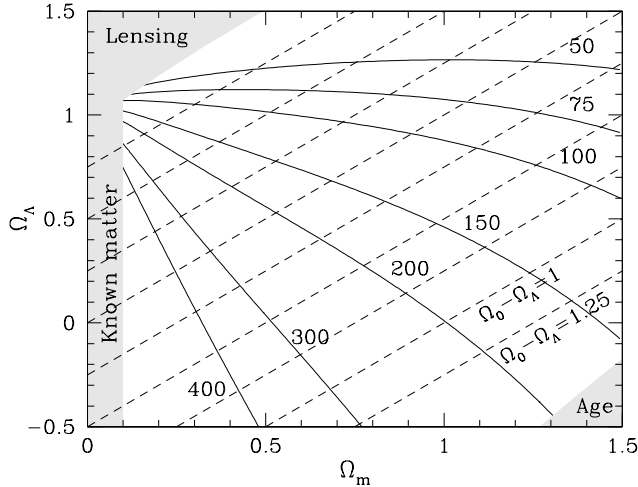}
\end{center}
\caption{The contours of constant $\Omega_0-\Omega_\Lambda$ (dashed lines,
in steps of 0.25) and constant $\ell_{\rm peak}$ in models with adiabatic
fluctuations, in the $\Omega_0$-$\Omega_\Lambda$ plane.
We have taken $h=0.65$ and $\Omega_{\rm B}h^2=0.02$ for definiteness.
There is a small, ${\cal O}(10\%)$, dependence of $\ell_{\rm peak}$ on
these values as discussed in the text.
Note that the contours are almost perpendicular in the observationally
preferred region of parameter space.  The shaded regions are ruled out by
other constraints: $\Omega_0<0.1$ is inconsistent with the amount of matter
observed, the lensing constraint is that the path length back to $z=2$ cannot
exceed 10 times the value in an Einstein-de Sitter universe and for the age
we have simply taken $H_0t_0>0.6$ as a lower limit.}
\label{fig:bigpicture}
\end{figure}

There are two groups which have published data on the redshift-luminosity
relation of SN-Ia at high-$z$, and both groups appear to favour a low
value of the mass density $\Omega_0$ (see later).
Additionally, the current results on CMB anisotropies provide a lower limit
on the angular scale of a peak in the spectrum.  Such a lower limit provides
an upper limit on the density of the universe in curvature, and thus a lower
limit on $\Omega_0$ in an open universe.
In this paper we combine the current data on temperature anisotropies with
the data on the SN-Ia redshift-distance relation.
The purpose is threefold.
The first and most important is to point out that the contours of constant
peak angular scale in the CMB and constant $q_0$ are nearly orthogonal in the
observationally favoured region of parameter space
(see Fig.~\ref{fig:bigpicture}).
The second is to see whether the current measurements, taken at face value,
are giving a consistent picture.
The third is to present the current status of the constraints, and indicate
how we expect them to change in the near future.

Of course we expect rapid progress in both of these fields, and we are aware
that a purely statistical analysis such as we present here can miss the
dominant sources of error.
However it seems timely to assess the level of agreement between these
very different techniques for determining $\Omega_0$ and $\Omega_\Lambda$.
The main conclusion of this work is that the two fields provide complementary
information, and that of course will be robust to changes in the data.

Finally a word about notation.
We assume the universe is described by a FRW metric with scale factor
$a(t)$ and critical density $\rho_{\rm crit}=3H_0^2/(8\pi G)$.
Here $H_0=100 h\,{\rm km}/{\rm s}/{\rm Mpc}$ is the Hubble constant.
We label the energy density, in units of $\rho_{\rm crit}$, from the various
species as: non-relativistic matter $\Omega_0$;
cosmological constant $\Omega_\Lambda$;
and curvature $\Omega_K$, where $K=-H_0^2(1-\Omega_0-\Omega_\Lambda)$.
These densities must sum to unity: $\Omega_0+\Omega_\Lambda+\Omega_K\equiv1$.
We define the dimensionless Hubble parameter $E(z)$ by
\begin{equation}
E^2(x=1+z) =
  \Omega_0 x^3+ \Omega_K x^2 + \Omega_\Lambda \qquad .
\end{equation}
In \S\ref{sec:xcdm} we will also discuss the recently popular idea of a flat
universe with a low matter density, the shortfall in energy density being
made up in a ``smooth'' component which we shall call $X$.  We shall see that
in this model the combination of peak location and SN-Ia distances is not as
powerful a constraint.

\section{Review}

Here we review some of the elementary material which will be used in this
paper.  This material is included for completeness and to define notation.
The reader familiar with this is urged to skip to the next section.

\subsection{Luminosity Distance}

The Type-Ia Supernovae results can be used to measure the luminosity
distance back to a given redshift $d_L(z)$.  The luminosity distance
is related to the coordinate distance by
\begin{equation}
d_L(z)={1+z\over\sqrt{|K|}}\ \sinh\left( \sqrt{|K|}\ r(z) \right)
\label{eqn:dL}
\end{equation}
for $K<0$.  For $K>0$ simply replace the $\sinh$ by a $\sin$.
At low redshift
\begin{equation}
H_0 r(z) = z\left( 1 + (1-q_0) {z\over 2} \right)\qquad z\ll 1
\end{equation}
where $q_0 = {1\over2}\Omega_0-\Omega_\Lambda$.
For the redshifts of relevance for the SN-Ia results however the low-$z$
expression is not accurate.  For arbitrary redshift
\begin{equation}
H_0 r(z) = \int_0^z {dz'\over E(z')}
\label{eqn:rz}
\end{equation}
which decreases to the present ($z=0$).

Measurements of the redshift-luminosity relation are usually presented in
terms of a distance modulus $m-M$ as a function of redshift.
With the definition of absolute magnitude, $M$, one has
\begin{eqnarray}
m - M &=& 5\log_{10}{d_L\over{\rm Mpc}} + 25 \\
&\simeq& 5\log_{10} {cz\over H_0} + \cdots\qquad {z\ll1}
\end{eqnarray}
which is dependent on the value of $h$ through the dependence in $d_L$.
The measured magnitude, $m$, is independent of $h$, but the absolute
magnitude, $M$, carries an $h$ dependence which cancels that of $d_L$.

For $z\ll 1$ the distance modulus depends on $\Omega_0$ and $\Omega_\Lambda$
only through the combination $q_0$, thus the slope of contours of constant
$m-M$ in the $\Omega_0$--$\Omega_\Lambda$ plane is ${1\over 2}$.
As one goes to higher $z$ one must use Eq.~(\ref{eqn:rz}) and the contours
are no longer straight lines in this plane.
However since the curvature is ``small'' we can associate an approximate
slope to these contours, which becomes steeper as $z$ is increased.
By $z=0.4$ the slope is approximately 1 over the range of $\Omega_0$ and
$\Omega_\Lambda$ of interest.  This increases to 1.2--1.3 by $z=0.6$,
1.5--2 by $z=0.8$ and 2--3 by $z=1$. (The range indicates the curvature
of the contours, which are steeper at higher $\Omega_0$.)
For the foreseable future most of the SN-Ia will be at $z\sim{1\over2}$,
so the likelihood contours will be elongated along
constant $\Omega_0-\Omega_\Lambda$.

\subsection{Angular Diameter Distance} \label{sec:peakloc}

The presence of any feature in the CMB anisotropy spectrum whose physical
scale is known provides us with the ability to perform the classical angular
diameter distance test at a redshift $z\simeq 10^3$
(see e.g.~Hu \& White~\cite{Acoustic}, Fig.~11).
Perhaps the most obvious feature, and probably the first whose angular
size will be measured, is the position of the first ``peak'' at degree
scales in the angular power spectrum.

For any model in which no ``new'' component contributes at $z\simeq 10^3$
the positions of the peaks (features) in $k$-space in the angular power
spectrum depend only on the physical densities of matter
$\omega_0\equiv\Omega_0 h^2$ and baryons
$\omega_{\rm B}\equiv\Omega_{\rm B} h^2$.
We assume that the CMB temperature is known well enough to be ``fixed'' and
that in addition to the CMB photons there are three massless neutrino species
contributing to the radiation density at $z\simeq 10^3$.

The peaks arise due to acoustic oscillations in the photon-baryon fluid
at last scattering, with photon pressure providing the restoring force,
baryons the fluid inertia and gravity the driving force
(e.g.~Bennett, Turner \& White~\cite{PhysTod} for an elementary review).
In adiabatic models, the assumption throughout this paper, the first peak
represents a mode which is maximally overdense when the universe recombines.
This peak is slightly broadened and shifted to larger angular scales by
contributions from the Integrated Sachs-Wolfe effect
(Sachs \& Wolfe~\cite{SacWol}) near last scattering
(Hu \& White~\cite{Acoustic}).

In models in which the fluctuations are not purely adiabatic, causality
requires that the features move to smaller angular scales, i.e.~higher
multipole moment $\ell$ (Hu, Spergel \& White~\cite{HSW}).
Thus in principle the measurements of a peak can provide only an {\it upper\/}
limit on $\Omega_K$.  Measurements of the curvature independent of the model
of structure formation require information on smaller scale anisotropy
(Hu \& White~\cite{TestInf,Damping}).
We shall henceforth assume that the fluctuations are adiabatic.  This can be
considered as adopting the upper limit on $\Omega_K$ as a measurement of
$\Omega_K$, which will be conservative for the purposes of this paper.
Alternatively one can argue that the adiabatic models are by far the best
motivated and additionally are observationally preferred, so we focus on
their predictions.
In the future, measurements of the whole anisotropy spectrum can allow us to
relax this constraint.
Under fairly reasonable assumptions (the fluctuations in the CMB were produced
by gravitational instability, the baryon content is constrained by
nucleosynthesis, secondary perturbations do not overwhelm the primary signal,
etc) these measurements can provide a model independent measurement of the
same combination of parameters as is constrained by $\ell_{\rm peak}$.
As CMB measurements become more precise we expect the ``error ellipse''
to be narrowed perpendicular to contours of constant $\ell_{\rm peak}$.
(The length of the ellipse along the contour depends on signals other than
the acoustic signature, which we will not consider here -- it is this
direction which is constrained by the SN.)

In order to proceed we have improved upon the approximation in
(Hu \& Sugiyama~\cite{Analytic}, White \& Scott~\cite{WhiSco})
by calibrating the position of the first peak in $k$-space by numerical
integration of the Einstein, fluid and Boltzmann equations.
A fit which is accurate to better than 1\% whenever $0.1\le\omega_0\le0.25$
and $0.01\le\omega_{\rm B}\le0.025$ is
\begin{equation}
\begin{array}{ll}
{\displaystyle k_{\rm peak}\over\displaystyle {\rm Mpc}^{-1} } \simeq &
  0.0112 + 0.0441 \omega_0 - 0.043\omega_0^2 \\
&  - 0.0496\omega_{\rm B} + 0.162\omega_0\omega_{\rm B}
   + 2.65\omega_{\rm B}^2 .
\end{array}
\end{equation}

Any feature, e.g.~$k_{\rm peak}$, then projects as an anisotropy onto an
angular scale associated with multipole
\begin{equation}
\ell_{\rm peak} \equiv k_{\rm peak} r_\theta
\end{equation}
where $r_\theta$ is the (comoving) angular diameter distance to last
scattering.  In terms of the conformal time $d\eta=dt/a(t)$ we have
\begin{equation}
  r_\theta = {1\over\sqrt{|K|}}
  \sinh\left[ \sqrt{|K|} \left(\eta_0-\eta_*\right) \right]
\label{eqn:rtheta}
\end{equation}
for $K<0$ negatively curved universes.  For $K>0$ merely replace the $\sinh$
with a $\sin$.  Note the similarity to the expression for $d_L(z)$ in
Eq.~(\ref{eqn:dL}).
Here $\eta_*(\Omega_0 h^2,\Omega_{\rm B} h^2)$ is the conformal time at
last scattering.
An accurate fit to the redshift of last scattering, $z_{\rm ls}\sim 10^3$,
can be found in Appendix E of (Hu \& Sugiyama~\cite{SmallScale}).
The conformal time at redshift $z$ is given by
\begin{equation}
H_0\eta(z) = \int_{z}^{\infty} {dz\over E(z)}
\end{equation} 
which increases to the present ($z=0$).
The accuracy of this projection for a range of models can be seen in
Hu \& White~(\cite{TestInf}), Fig.~2, where it is seen to be good to
${\cal O}(15\%)$ except near the edges of parameter space.  Slightly
better agreement is obtained for the higher peaks or peak spacings,
though we shall consider only the first peak here.
In the figures we have corrected for the ${\cal O}(15\%)$ error induced
by this approximation, using the numerical computation of the anisotropy
spectrum.

Note that since the anisotropy is being produced at $z\gg 1$, the contours
have rotated to be almost orthogonal to those probed by the SN-Ia.
The cosmological constant is dominating only at late times, which make up
most of the range of redshift probed by the SN-Ia.  The conformal age today
however probes most of the matter dominated epoch and thus has a very
different dependence on the density parameter and cosomological constant.

\section{The Data}

In this section we describe the statistical analysis of the existing data.
We are of course aware that statistical uncertainties are not the only,
and perhaps not the major, uncertainty in all of the data sets at this early
stage.  Of particular concern for SN-Ia are the effects of extinction in the
host galaxy (but see Branch~\cite{Branch}) or differential reddening
corrections between the distant and local/calibrating samples.  For the CMB
the difficulties associated with calibration and foreground extraction are
the most worrisome.
Here we shall describe our treatment of the statistical problem and leave
the question of systematics to be determined by further data
(see also Riess et al.~\cite{HighZnew}).

\subsection{Other Constraints} \label{sec:prior}

Let us first start by outlining the situation before the constraints from
SN-Ia and the CMB are imposed (see also White \& Scott~\cite{WhiSco}).
The amount of matter in the universe inferred from dynamical measurements
gives a lower bound on $\Omega_0\ga0.1$ which rules out the region on the
far left of our figures.
The relative scarcity of gravitational lenses gives a {\it conservative\/}
bound that the path length back to redshift $z\sim 2$ be $\la10$ times the
Einstein-de Sitter value.  This rules out the upper-left corner, which would
anyway be ruled out by requiring the presence of a big bang or by the
existence of high-$z$ objects (Caroll et al.~\cite{CPT}).
We have imposed a constraint on the lower right of the figure that the age
of the universe $H_0t_0>0.6$, which corresponds to 12Gyr for $h=0.5$.
This is less conservative than the other limits, but that region is not
observationally preferred anyway.

A model dependent upper limit to $\Omega_0+\Omega_\Lambda$ comes from the
{\sl COBE\/} data, which constrain $\Omega_0+\Omega_\Lambda\la1.5$ under
reasonable assumptions about the initial power spectrum of fluctuations
(White \& Scott~\cite{WhiSco}).  We have not shown this region as shaded
because of the model dependence of the constraint.
Further discussion of constraints can be found in
White \& Scott~(\cite{WhiSco}).

\subsection{Supernovae}

Two teams have been finding high-$z$ Type-Ia supernovae suitable for
measuring the $d_L(z)$ relation.
The `Supernova Cosmology Project'\footnote{http://www-supernova.lbl.gov/}
has published data on 8 supernovae, of which 6 can be corrected for the
width luminosity relation (Perlmutter et al.~\cite{Perl1,Perl2}).
The `High-z Supernova
Search'\footnote{http://cfa-www.harvard.edu/cfa/oir/Research/
supernova/HighZ.html}
has published data on 10 supernovae (Garnovich et al~\cite{HighZ},
Riess et al.~\cite{HighZnew}).
In addition a set of low-$z$ `calibrating' supernovae has been obtained
by the Calan-Tololo group (Hamuy et al.~\cite{CalTol}).
We use the 26 supernovae with $B-V<0.2$ described in
Hamuy et al.~(\cite{CalTol}).

\begin{figure}[t]
\begin{center}
\leavevmode
\epsfxsize=3.4in \epsfbox{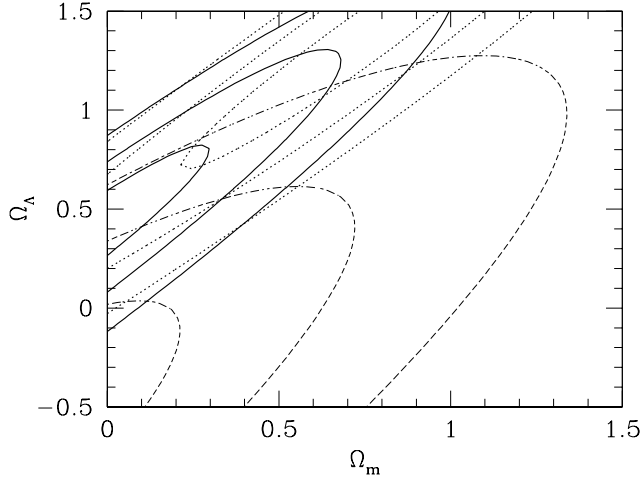}
\end{center}
\caption{Likelihood in the $\Omega_0$-$\Omega_\Lambda$ plane for fitting
the SN-Ia data described in the text.  The dashed lines are $1\sigma$,
$2\sigma$ and $3\sigma$ contours for fitting the low-$z$ results plus the
Supernova Cosmology Project data.  The dotted lines are for fitting the
low-$z$ plus the High-Z data and the solid lines are for fitting all of
the data.}
\label{fig:sn1}
\end{figure}

For simplicity we use only the $B$ band observations for all of the
supernovae.
For the Hamuy et al.~(\cite{CalTol}) sample, the observational errors on
$\log_{10}(cz)$ were added in quadrature to the quoted errors on the
(uncorrected) peak magnitude using the fact that at low-$z$ the slope of
the magnitude $\log_{10}(cz)$ relation is 5.
We performed a maxmimum likelihood fit to the combined data set as follows.
First for a given point ($\Omega_0,\Omega_\Lambda$) we computed the expected
function $m_B(z)$, with a given absolute brightness $M_B$ of SN-Ia.
Each of the data points was then width-luminosity corrected using a linear
relation (Hamuy et al.~\cite{CalTol})
\begin{equation}
  m_{{\rm corr},B} = m_B - \alpha\left(\Delta m_{15}-1.1\right)
\label{eqn:wlum}
\end{equation}
where $\Delta m_{15}$ is the number of magnitudes the SN-Ia declines in the
first 15 days (Phillips~\cite{Phillips}; for several of the Calan-Tololo
SN-Ia the lightcurves do not start until 10 days after the peak, in this case
$\Delta m_{15}$ is obtained from a fit to a template SN-Ia in the BV\&I bands).
For fixed $\alpha$ the errors on $\Delta m_{15}$ and $m_B$ were added in
quadrature, ignoring a potential for correlated errors from photometric
uncertainties affecting the lightcurve fit.  These correlations are not quoted
by any of the groups, though they are included in the fit by the Supernova
Cosmology Project team (Perlmutter, private communication).
Including such correlations would only serve to tighten the allowed regions,
so neglecting them is conservative.
Performing a $\chi^2$ fit of the data to the theory then allows us to
calculate
\begin{equation}
{\cal L}(\Omega_0,\Omega_\Lambda,M_B,\alpha) =
  \exp\left[ -{\chi^2\over 2} \right] \qquad .
\end{equation}
This likelihood is then marginalized over $M_B$ with a uniform prior and
over $\alpha=0.784\pm0.182$ (Hamuy et al.~\cite{CalTol}) where the error
is assumed to be gaussian.
The result is ${\cal L}(\Omega_0,\Omega_\Lambda)$ which is shown in
Fig.~\ref{fig:sn1}.

Note that this method is not exactly equivalent to the analysis performed by
either of the groups whose supernovae data we have used.
In particular we have used only a subset of the data (the $B$-band), we have
used Eq.~(\ref{eqn:wlum}) rather than a multi-color lightcurve shape method
and we have marginalized over the absolute brightness of SN-Ia and slope of
the width-luminosity relation Eq.~(\ref{eqn:wlum}).
By marginalizing over $\alpha$ and $M_B$ the fit is allowed to find the best
value for the combined data set, rather than fixing e.g.~$M_B$ from the
low-$z$ sample alone.
We find that the maximum likelihood point
$(\Omega_0,\Omega_\Lambda,M_B,\alpha)$ is a reasonable fit to the data,
being allowed at 96\% confidence level.
The subset excluding the High-Z data fares better: it is allowed at 78\% CL.
The subset excluding the SNCP data is allowed at the 93\%CL.
The best fit to all of the data has $M_B$ less than $1\sigma$ from the best
fit to the Hamuy et al.~(\cite{CalTol}) data though before marginalization
prefers a higher value of $\alpha\simeq1.1$.  Again excluding the High-Z
data the best fit prefers $\alpha\simeq0.784$ in agreement with
Hamuy et al.~(\cite{CalTol}).  Given all this, the results of the
marginalization procedure should be very close to the alternative method of
maximizing ${\cal L}(\Omega_0,\Omega_\Lambda)$ over $\alpha, M_B$.  Indeed
the likelihood contours derived here compare well with those in
Perlmutter et al.~(\cite{Perl2}) and Riess et al.~(\cite{HighZnew}).

\subsection{Anisotropy}

The observational situation with regards anisotropy measurements on degree
scales, which can pin down the location of the first peak in the spectrum,
is not as advanced as for the supernovae.
It is common procedure to constrain cosmological models (usually variants of
cold dark matter, a.k.a.~CDM) by fitting theories to a collection of
``bandpowers''.
For current data this provides a good approximation to the results of a full
likelihood analysis (Bond, Jaffe \& Knox~\cite{BonJafKno}).
Recent work using this method gives $\ell_{\rm peak}\sim260$
(Lineweaver~\cite{Charley}) or $\ell_{\rm peak}=263^{+139}_{-94}$
(Hancock et al.~\cite{HanRocLasGut}; the error is almost symmetric in
$\log\ell$).
This would constrain $\Omega_0$ in an open universe to $\Omega_0\ga0.4$
(95\%CL, Lineweaver~\cite{Charley}; Hancock et al.~\cite{HanRocLasGut}).

Some of the weight against low-$\Omega_0$ models in the above fits comes
from the fact that low-$\Omega_0$ CDM models typically predict a ``dip''
in power before the rise into the first peak.
This dip comes about because the ISW effect provides an increase in
large-angle power while both the ISW effect and the acoustic oscillations
in the baryon-photon fluid provide a rapid rise in power into the first peak.
This leads to a ``dip'' in power just before the rise into the first peak
which is constrained by the present data.
However, since the behaviour on angular scales larger than the peak can be
model dependent we must be careful in applying the limits quoted above.

\begin{figure}[t]
\begin{center}
\leavevmode
\epsfxsize=3.4in \epsfbox{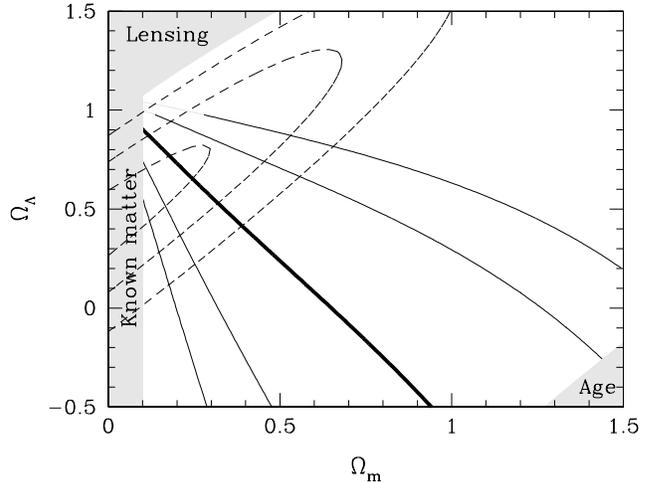}
\end{center}
\caption{Likelihood in the $\Omega_0$-$\Omega_\Lambda$ plane for fitting
both the SN-Ia and CMB data described in the text.
The dashed lines are $1\sigma$, $2\sigma$ and $3\sigma$ contours for fitting
the SN-Ia results, the solid lines the CMB results.  The thick solid contour
represents the peak of the likelihood found by
Hancock et al.~(\protect\cite{HanRocLasGut}), and the two contours to either
side represent conservative $\pm1\sigma$ and $\pm2\sigma$ values.
The shaded areas are as in Fig.~\protect\ref{fig:bigpicture}.}
\label{fig:both}
\end{figure}

Perhaps the most conservative estimate of the location thus comes from
the analysis of Hancock et al.~(\cite{HanRocLasGut}), who in addition
to CDM models fit a phenomenological model first proposed in
(Scott, Silk \& White~\cite{SSW}).  This phenomenological model doesn't
contain the ``dip'' and thus penalizes the high-$\ell_{\rm peak}$ models
based only on the information around the peak.  It is on the basis of this
model that they quote a marginalized peak position
$\ell_{\rm peak}=263^{+139}_{-94}$ which we have used in Fig.~\ref{fig:both}.

\subsection{Combined Constraint}

The allowed regions from the SN-Ia, CMB peak and constraints discussed in
\S\ref{sec:prior} are shown in Fig.~\ref{fig:both}.
Since the current CMB data do not unambiguously show a well defined peak we
have been cautious in plotting the allowed CMB region.  The contours shown
cover the region $\ell_{\rm peak}\la 550$ (lower left) to
$\ell_{\rm peak}\ga 130$ (upper right).  This corresponds approximately to
a 95\% CL.  The $2\sigma$ limit at low-$\ell$ is stronger than twice the
$1\sigma$ error quoted above since the likelihood function is non-gaussian
(Hancock et al.~\cite{HanRocLasGut}).
The upper limit at $\ell\simeq550$ is probably less firm than the lower
limit, for which there is more data, and is quite sensitive to the results of
the CAT experiment (Scott et al.~\cite{CAT}).
For a given (e.g.~CDM) model it is possible to obtain stronger constraints
on the parameters $\Omega_0$ and $\Omega_\Lambda$, but even the model
independent constraint is quite promising.

\section{Smooth Component} \label{sec:xcdm}

\begin{figure}[t]
\begin{center}
\leavevmode
\epsfxsize=3.4in \epsfbox{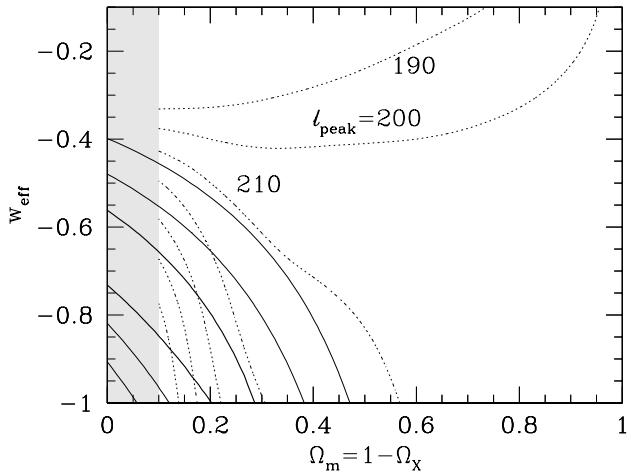}
\end{center}
\caption{Likelihood in the $\Omega_0$-$w$ plane for fitting
the SN-Ia data described in the text.  Solid lines are $1\sigma$,
$2\sigma$ and $3\sigma$ contours.  The dotted lines represent contours
of constant $\ell_{\rm peak}$ for $h=0.65$ and $\Omega_{\rm B}h^2=0.02$
in steps of $\Delta\ell=10$.}
\label{fig:xcdm}
\end{figure}

The discussion until now has been focussed on the ``traditional'' cosmological
parameters $\Omega_0$ and $\Omega_\Lambda$.  However it has recently become
fashionable to postulate the existence of a heretofore unknown component whose
contribution to the energy density makes $\Omega_{\rm tot}=1$ or equivalently
$\Omega_K=0$.
Here we shall refer to this component as $X$, indicating its unknown nature.
For the purposes of this paper all we shall need to know about $X$CDM is that
it is smooth on scales smaller than the horizon and its equation of state is
$w\equiv p_X/\rho_X$ which we shall further assume is a constant.
We will require that $w<0$ so that $X$-matter is only contributing appreciably
to the energy density at low-$z$.
$X$-matter will serve here as an example of a different set of distance
redshift relations which can be probed by the combination of SN-Ia and CMB
anisotropy measurements.

Various examples of $X$CDM, plus references, are given in
Turner \& White~(\cite{xCDM}).
A particularly appealing example of $X$-matter is a scalar field, which has
been treated in detail recently by
(Coble, Dodelson \& Frieman~\cite{CDF} and
Caldwell, Dave \& Steinhardt~\cite{QCDM};
see also Ferreira \& Joyce~\cite{FerJoy})
though the idea goes back much further (e.g.~Kodama \& Sasaki~\cite{KodSas}).
In this case there is the probability that $w$ will vary with time, which
will modify the results presented below as discussed later.

We also comment briefly on more exotic possibilities.
When adding a ``new'' component the behaviour of the background space-time
and the perturbations are governed by the stress-energy tensor of the new
component(s).  In Turner \& White~(\cite{xCDM}) the new component was
agnostically approximated as perfectly smooth, i.e.~the perturbations in the
stress-energy tensor were neglected.
For a scalar field this is quite a good approximation well below the horizon,
though it has been noted (Caldwell et al.~\cite{QCDM}) that such a
prescription is gauge dependent and thus unphysical on scales larger than
the horizon.
If one assumes that $X$-matter is a scalar field then in addition to the
``smooth'' component one needs only specify one fluctuating component: the
energy density (which is related to the pressure and velocity fluctuations).
All other fluctuations are zero.
This can be generalized by allowing for a non-vanishing anisotropic stress
(or ``viscosity'') which leads to ``generalized'' dark matter
(Hu~\cite{GDM}) which has many nice properties.  This model is called
``generalized'' because specifying the density, velocity, pressure and
anisotropic stress of the new component completely specifies the scalar
part of the stress energy tensor.
Of course beyond ``generalized'' dark matter one can include not only scalar
(density) perturbations but vector and tensor perturbations, and multi-field
models.  All of these possible additions will have observable effects on the
CMB anisotropy spectrum and large-scale structure.
However, the parameter space is so large that we shall concentrate simply on
$X$-matter with constant $w$.
Under this assumption $\rho_X\sim a^{-3(1+w)}$.  Two familiar limits are
$w=-1$ for $\Omega_\Lambda$ and $w=-1/3$ for $\Omega_K$.

Assuming that the vector and tensor components, anisotropic stress and
2-field isocurvature component are negligible, the only effect of $X$-matter
on the CMB primary anisotropy spectrum is to change $r_\theta$
(Eq.~\ref{eqn:rtheta}) and modify the large-angle anisotropies.
One can calculate $\ell_{\rm peak}$ merely by generalizing $E^2(x=1+z)$ to
include a component $\Omega_X x^{3(1+w)}$ in the calculation of $r_\theta$.
The redshift-luminosity distance is only affected by the background energy
density contributed by $X$-matter, which again is completely encoded by
$E(z)$.

The situation with respect to large-angle anisotropies (e.g.~as probed by
{\sl COBE\/}) is more complex.
The simplistic assumption that $X$-matter be completely smooth can not be
exact on length scales approaching the horizon.  A calculable example is
provided by scalar field models.
As stressed by Caldwell et al.~(\cite{QCDM}), though a scalar field is
{\it approximately\/} smooth the deviations from this can affect large-angle
anisotropies, and hence the {\sl COBE\/} normalization
(Bennett et al.~\cite{COBE}, Hinshaw et al.~\cite{Hinshaw},
Gorski et al.~\cite{Gorski}, Banday et al.~\cite{Banday};
we have used the method of Bunn \& White~(\cite{BunWhi}) to normalize these
models).
Indeed, comparing the results of Turner \& White~(\cite{xCDM}) with a
calculation of the scalar field case with constant $w$, we find that the
fluctuations can give up to a $30\%$ correction to the large-scale
normalization in the range $0.2<\Omega_0\le 1$ and $-1\le w< -1/3$ with the
trend being worse agreement for higher $w$ and lower $\Omega_0$.
For $w\la -2/3$ or $\Omega_0\ga0.4$ the correction is negligible
(the data currently prefer $w\approx -1$).
Unlike the case of precisely smooth $X$-matter, the scalar field matter power
spectra are not very well approximated by the oft-used $\Gamma$-models
(Bardeen et al.~\cite{BBKS}) once $w\ne -1$ (Hu~\cite{GDM}, Fig.~4).
The sense is to reduce the ratio of small- to large-scale power as $w$ is
increased (i.e.~made less negative) with the decrease occuring around the
sound horizon at scalar field domination.
For $w\la-2/3$ and $\Omega_0\ga0.3$ the correction is $\la20\%$, but for
larger $w$ can be a factor of 5.  This makes the extrapolation to small
scales more model dependent.

Unfortunately if we believe that $X$-matter is a scalar field, the motivation
for assuming a constant $w$ is somewhat weak.  While we can define an
``average'' value of $w$ to use for calculating $\ell_{\rm peak}$, the effect
on the large-angle anisotropies is more complex.
The situation for ``generalized'' dark matter, cosmic strings and textures
is even more uncertain.  Here anisotropic stress can play a large role, as
pointed out by Hu~(\cite{GDM}), and lead to potentially large modifications
of the large-angle anisotropies.  Thus for these models the {\sl COBE\/}
normalization should be treated as highly uncertain -- though the trend is
to decrease the normalization over the $w=-1$ case.
For general $X$-matter the best normalization currently comes from the local
abundance of rich clusters -- see Wang \& Steinhardt~(\cite{WanSte}) for a
discussion.  However without an independent normalization to compare to,
this cannot be considered as a strong constraint on the model.

For these reasons we will focus on smaller angular scales, where the
anisotropies were generated before $X$-matter became dynamically important.
Specifically we shall focus on $\ell_{\rm peak}$, and treat $w$ as an
appropriate average value occuring in $E(z)$.
If we ignore the large-angle ISW effect (which can anyway be obscured by
cosmic variance) and the effects of gravitational lensing on the small-scale
anisotropy, then models which hold $\eta_0$ and the physical matter and baryon
densities fixed will result in degenerate spectra.  For our purposes this
means they will predict the same $\ell_{\rm peak}$, however we choose to
measure it.  Hence $X$-matter introduces only one extra parameter (not time
varying function) to be constrained by $\ell_{\rm peak}$.
Unfortunately, while $\ell_{\rm peak}$ in insensitive to the details of
$X$-matter, because we have kept the spatial hypersurfaces flat it also
does not show the large variations seen in the $\Omega_0$--$\Omega_\Lambda$
plane.
Given the embryonic state of the CMB peak measurements to date, the peak
location provides little constraint.
Thus $X$-matter serves as an example where the combination of CMB and
SN-Ia results does not highly constrain the parameter space (as yet),
in contrast to the standard $\Omega_0$--$\Omega_\Lambda$ case.

We show in Fig.~\ref{fig:xcdm} the 1, 2 and $3\sigma$ contours in the
$\Omega_X$--$w$ plane from the full SN-Ia data set.  This updates Fig.~4
of Turner \& White~(\cite{xCDM}).  Contours of $\ell_{\rm peak}$ are
superposed on Fig.~\ref{fig:xcdm} to show the predicted location of the
peak is quite insensitive to changes in $\Omega_X$.  These contours are
for $h=0.65$ and $\Omega_{\rm B}h^2=0.02$, changing these parameters can
affect $\ell_{\rm peak}$ by ${\cal O}(10\%)$.
As discussed above the approximation discussed in \S\ref{sec:peakloc} is
only expected to work to ${\cal O}(15\%)$, which is about the amount by which
$\ell_{\rm peak}$ changes in the $X$-matter case.  Thus we have used a full
numerical computation of $\ell_{\rm peak}$ in making this figure.

As can be seen the region near $\Omega_0=0.3$ is allowed for $w\approx -1$,
in agreement with earlier studies
(Coble, Dodelson \& Frieman~\cite{CDF}; Turner \& White~\cite{xCDM};
Caldwell et al.~\cite{QCDM}).  Adding additional constraints such as
the shape of the matter power spectrum as measured by galaxy clustering,
high-$z$ object abundances and the abundance of rich clusters today does
not alter this conclusion.
While enhanced SN-Ia measurements will pick out a strip in the $\Omega_0$--$w$
plane, strong constraints on $X$-matter scenarios will probably have to await
the upcoming satellite CMB anisotropy missions, especially if $w$ changes at
high-$z$.  Even with CMB satellites, accurate CMB measurements will have
difficulty in determining $\Omega_X$ and $w_{\rm eff}$ precisely and
simultaneously unless they can probe high enough $\ell$ to see the effects
of gravitational lensing.
(Lensing depends on the matter power spectrum, which probes a different
combination of parameters than $\eta_0(\Omega_X,w)$=const, breaking the
$\Omega-w$ degeneracy which enhances the marginalized errors.)
Thus the combination of CMB and SN-Ia results will be important even in
the ``{\sl MAP\/}\footnote{http://map.gsfc.nasa.gov/} era''.

\section{Conclusions and the Future}

We have examined the constraints in the $\Omega_0$--$\Omega_\Lambda$ plane
arising from a combination of SN-Ia and CMB data.  Our most important result
is that the two data sets provide approximately orthogonal constraints and
thus nearly maximal complementarity.  We have illustrated this by a likelihood
analysis of the current data.
Even though very different uncertainties affect the two data sets, the
likelihood functions are compatible, with the allowed region shown in
Fig.~\ref{fig:both}.  As more data is acquired, the combination will serve
as an important cross check.

Due to its current fashionability, and for contrast, we have also looked at
constraints on universes with flat spatial hypersurfaces and low-$\Omega_0$,
with $1-\Omega_0$ in a smooth component called $X$.
As expected the constraints on this model are much weaker since -- by
design -- the CMB peak location varies little with the model parameters.
The allowed region is near $w\approx -1$ and the CMB peak does not prefer
any value of $\Omega_0$ within this region.

What about the future of this enterprise?  Both the Supernova Cosmology
Project and the High-Z team have $\sim50$ supernovae still to be analyzed,
which should shrink the contours in Fig.~\ref{fig:sn1} considerably.
Multicolour data will help control the uncertainty due to redenning and
the allowed region should lie along a line of slope $\simeq1$ (most SN-Ia
will be at $z\sim{1\over2}$) in the $\Omega_0$--$\Omega_\Lambda$ plane with
width $\pm0.1$.
On the CMB front, data from currently operational experiments could
determine the location of the peak in $\ell$ to $\Delta\ell\la 30$
within the next year (assuming the peak is near $\ell\simeq 250$).
{}From Fig.~\ref{fig:bigpicture} we see that such a measurement of
$\ell_{\rm peak}$ would be comparable to, but orthogonal to, the supernova
constraint.
The first such experiment, the Mobile Anisotropy
Telescope\footnote{http://dept.physics.upenn.edu/$\sim$www/astro-cosmo/
devlin/project.html} has already had a ``season'' in Chile.
The Viper\footnote{http://cmbr.phys.cmu.edu/vip.html} telescope is operating
at the South Pole and results from several other experiments
(see Bennett et al.~\cite{PhysTod}; Table~1) are expected this year or next.
And of course we anticipate that {\sl MAP}, scheduled for launch in late 2000,
will determine $\ell_{\rm peak}$ to a few per cent, the precise number
depending on the angular scale of the peak.

\bigskip
% \acknowledgments  
I would like to thank Joanne Cohn for a careful reading of the manuscript,
Wayne Hu for useful conversations on ``generalized'' dark matter and scalar
fields, Bob Kirshner and Saul Perlmutter for discussions on the supernova
constraints, Douglas Scott for comments on the manuscript and
Limin Wang for communicating his work on the cluster abundance ahead of
publication.


\begin{thebibliography}{99}
\bibitem[1997]{Banday}
Banday A., et al., 1997, \apj, { 475}, 393
\bibitem[1986]{BBKS}
Bardeen J.M., Bond J.R., Kaiser N., Szalay A.S., 1986, \apj, { 304}, 15
\bibitem[1996]{COBE}
Bennett C.L., et al., 1996, \apj, { 464}, L1
\bibitem[1997]{PhysTod}
Bennett C.L., Turner M.S., White M., 1997, Physics Today, November, 32.
\bibitem[1998]{BonJafKno}
Bond J.R., Jaffe A., Knox L., 1998, preprint [astro-ph/9708203]
\bibitem[1997]{Branch}
Branch D., 1997, \mnras, 290, 360
\bibitem[1997]{BunWhi}
Bunn E.F. \& White M., 1997, \apj, { 480}, 6
\bibitem[1992]{CPT}
Carroll S.M., Press W.H., Turner E.L., 1992, Ann. Rev. Astron.\& Astrophys.,
  { 30}, 499.
\bibitem[1997]{CDF}
Coble K., Dodelson S., Frieman J., 1997, Phys. Rev. { D55}, 1851
\bibitem[1998]{QCDM}
Caldwell R.R., Dave R., Steinhardt P.J., Phys. Rev. Lett, in press
  [astro-ph/9708069]
\bibitem[1997]{FerJoy}
Ferreira P.G., Joyce M., 1997, Phys. Rev. Lett., { 79}, 4740
\bibitem[1997]{HighZ}
Garnavich P.M., et al., 1997, preprint [astro-ph/9710123]
\bibitem[1995]{GooPer}
Goobar A., Perlmutter S., 1995, \apj, { 450}, 14
\bibitem[1996]{Gorski}
Gorski K., et al., 1996, \apj, { 464}, L11
\bibitem[1996]{CalTol}
Hamuy M., et al., 1996, \aj, { 112}, 2398
\bibitem[1997]{HanRocLasGut}
Hancock S., Rocha G., Lasenby A., Gutierrez C.M., \mnras, in press
  [astro-ph/9708254]
\bibitem[1996]{Hinshaw}
Hinshaw G., et al., 1996, \apj, { 464}, L17
\bibitem[1998]{GDM}
Hu W., 1998, \apj, in press [astro-ph/9801234]
\bibitem[1997]{HSW}
Hu W., Spergel D.N., White M., 1997, Phys. Rev., { D55}, 3288
\bibitem[1995]{Analytic}
Hu W., Sugiyama N., 1995, \apj, { 444}, 489
\bibitem[1996]{SmallScale}
Hu W., Sugiyama N., 1996, \apj, { 471}, 542
\bibitem[1996a]{Acoustic}
Hu W., White M., 1996a, \apj, { 471}, 30
\bibitem[1996b]{TestInf}
Hu W., White M., 1996b, Phys. Rev. Lett., { 77}, 1687
\bibitem[1997]{Damping}
Hu W., White M., 1997, \apj, { 479}, 568
\bibitem[1984]{KodSas}
Kodama H., Sasaki M., 1984, Prog. Theor. Phys., { 78}, 1
\bibitem[1998]{Charley}
Lineweaver C.H., 1998, in Proceedings of the Kyoto IAU Symposium 183:
``Cosmological Parameters and the Evolution of the Universe'', 
Kyoto, Japan, August 1997, Kluwer, in press [astro-ph/9801029]
\bibitem[1997]{Perl1}
Perlmutter S., et al., 1997, \apj, { 483}, 565
\bibitem[1998]{Perl2}
Perlmutter S., et al., 1998, Nature, { 391}, 51 [astro-ph/9712212]
\bibitem[1994]{Phillips}
Phillips M.M., 1993, \apj, { 413}, L105
\bibitem[1998]{HighZnew}
Riess A., et al., preprint [astro-ph/9805201]
\bibitem[1967]{SacWol}
Sachs R.K., Wolfe A.M., 1967, \apj, { 147}, 73
\bibitem[1995]{SSW}
Scott D., Silk J., White M., 1995, Science, { 268}, 829,
\bibitem[1996]{CAT}
Scott P.F.S.,  et al., 1996, \apj, { 461}, L1
\bibitem[1997]{xCDM}
Turner M.S., White M., 1997, Phys. Rev. D56, 4439
\bibitem[1998]{WanSte}
Wang L., Steinhardt P., preprint [astro-ph/9804015]
%\bibitem[1997]{SWsimple}
%White M., Hu W., 1997, Astron. \& Astrophys., { 321}, 8W
\bibitem[1996]{WhiSco}
White M., Scott D., 1996, \apj, { 459}, 415
\end{thebibliography}
\end{document}